\documentclass[aps,prd,twocolumn,preprintnumbers,superscriptaddress,floatfix]{revtex4}

\usepackage{multirow, graphicx,amssymb,url,mathrsfs,amsmath}
\usepackage{eucal,wrapfig,boxedminipage,setspace,subfigure}
\usepackage{amsxtra,amstext,latexsym,dsfont}

        \def\k  {\kappa}
             
          \def\s  {\sigma}

\def\IR{{\hbox{{\rm I}\kern-.2em\hbox{\rm R}}}}
\def\IB{{\hbox{{\rm I}\kern-.2em\hbox{\rm B}}}}
\def\IN{{\hbox{{\rm I}\kern-.2em\hbox{\rm N}}}}
\def\IC{\,\,{\hbox{{\rm I}\kern-.59em\hbox{\bf C}}}}
\def\IZ{{\hbox{{\rm Z}\kern-.4em\hbox{\rm Z}}}}
\def\IP{{\hbox{{\rm I}\kern-.2em\hbox{\rm P}}}}
\def\IH{{\hbox{{\rm I}\kern-.4em\hbox{\rm H}}}}
\def\ID{{\hbox{{\rm I}\kern-.2em\hbox{\rm D}}}}

\newcommand{\beq}{\begin{equation}}
\newcommand{\eeq}{\end{equation}}
\newcommand{\bea}{\begin{eqnarray}}
\newcommand{\eea}{\end{eqnarray}}

\begin{document}

\voffset 1cm

\newcommand\sect[1]{\emph{#1}---}

\title{Inverse Magnetic Catalysis in Bottom-Up Holographic QCD}

\author{Nick Evans, Carlisson Miller \&  Marc Scott}
\affiliation{STAG Research Centre \&  Physics and Astronomy, University of
Southampton, Southampton, SO17 1BJ, UK}

\begin{abstract}
We explore the effect of magnetic field on chiral condensation in QCD via a simple bottom up holographic model which inputs QCD dynamics through the running of the anomalous dimension of the quark bilinear. Bottom up holography is a form of effective field theory and we use it to explore the dependence on the coefficients of the two lowest order terms linking the magnetic field and the quark condensate. In the massless theory, we identify a region of parameter space where magnetic catalysis occurs at zero temperature but inverse magnetic catalysis at temperatures of order the thermal phase transition. The model shows similar non-monotonic behaviour in the condensate with B at intermediate T as the lattice data. This behaviour is due to the separation of the meson melting and chiral transitions in the holographic framework.  The introduction of quark mass raises the scale of B where inverse catalysis takes over from catalysis until the inverse catalysis lies outside the regime of validity of the effective description leaving just catalysis.  \noindent

\end{abstract}

\maketitle

\section{Introduction}

The study of strongly coupled theories at finite temperature in the presence of an external magnetic field is a topic of great interest for QCD. Cosmologically,  large magnetic fields may have been present at phase transitions \cite{Vachaspati:1991nm,Enqvist:1993np} and such conditions are also being produced in collisions at the Large Hadron Collider (LHC) and Relativistic Heavy Ion Collider (RHIC) \cite{Skokov:2009qp}.  A key question is how they impact on the thermal, chiral restoration transition. At zero temperature the strong dynamics of QCD forms a non-zero chiral condensate that breaks the gloabl chiral flavour symmetries to the vector subgroup. At high temperatures where asymptotic freedom sets in and renders the coupling small the condensate vanishes. The two phases are separated by a second order transition \cite{Ejiri:2009,Aoki:2006}.

\begin{figure}[]
 \centering 
{\includegraphics[width=8.5cm]{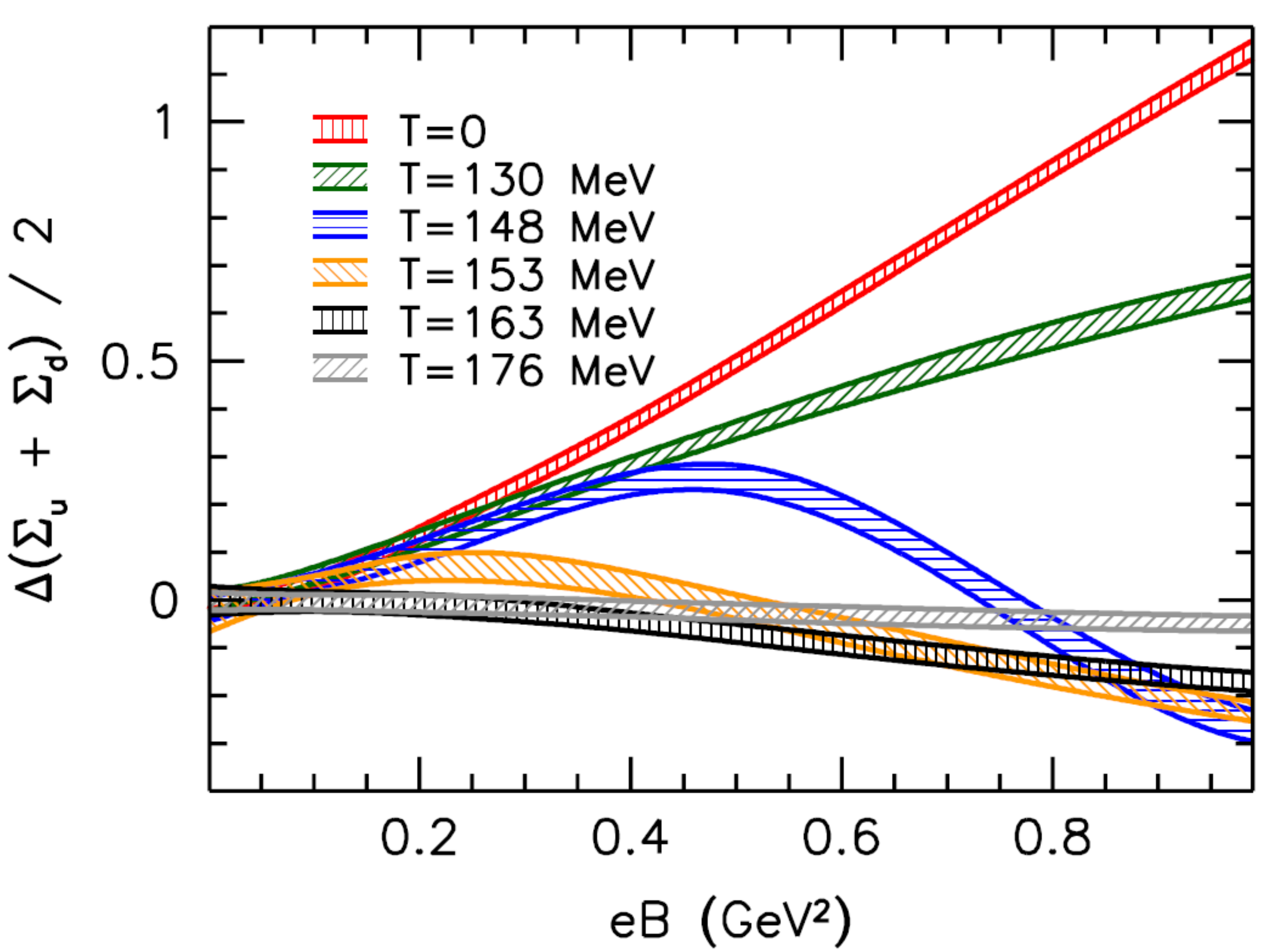}   }
  \caption{ Lattice results for the change in the quark condensate as a function of magnetic field strength over a range of temperatures - figure taken from \cite{Bali:2012zg}. }
            \label{fig:lattice}
\end{figure}

Recent lattice studies of QCD with light quarks and an applied magnetic field \cite{Bali:2011qj,Bali:2012zg,Endrodi:2015oba} have revealed some surprisingly complex behaviour. At zero temperature the magnetic field enhances the chiral condensate  $\s\equiv\langle\bar{q}q\rangle$ - ``magnetic catalysis'' -  as has generally been predicted \cite{Fraga:2008um,Mizher:2010zb,Gatto:2010pt,Kharzeev:2013jha,Miransky:2015ava}. At temperatures near the critical temperature though the B field has been shown to reduce the chiral condensate and reduce the critical temperature of the transition - ``inverse magnetic catalysis''. At intermediate temperatures there is a non-monotonic behaviour with small $B$ favouring chiral condensation but larger $B$ disfavouring it. These results are summarized in Fig  \ref{fig:lattice} taken from \cite{Bali:2012zg}. 

There has been considerable work in a number of approaches to explain these results %\cite{Fraga:2012ev}
\cite{Fraga:2012ev,Fukushima:2012kc,Kojo:2012js,Endrodi:2013cs,Bruckmann:2013oba,Chao:2013qpa,Ferreira:2013tba,Andersen:2013swa,Mueller:2014tea,Ruggieri:2014bqa,Farias:2014eca,Ferreira:2014kpa,Ayala:2014iba,Ayala:2014gwa,Ferrer:2014qka,Yu:2014xoa,Braun:2014fua,Mueller:2015fka,Ayala:2015lta,Ayala:2015bgv,Hattori:2015aki,Ahmad:2016iez,Mao:2016fha}. One such approach is holography  \cite{Maldacena,Gubser,Witten,Aharony:1999ti,Erdmenger:2007cm}
which we will explore further here. In principle there should be a stringy holographic dual of QCD but it is not known and may be strongly coupled itself. Top-down duals of confining gauge dynamics, including those with quark fields, although typically in the quenched approximation, do exist and show many similar behaviours to QCD \cite{Erdmenger:2007cm,Karch:2002sh,Grana:2001xn,Bertolini:2001qa,Kruczenski:2003be,Babington:2003vm}. These models have inspired bottom-up phenomenological models using the holographic framework that also provide a decent description of the QCD meson spectra \cite{Erlich:2005qh, DaRold:2005zs} and the quark-gluon plasma \cite{CasalderreySolana:2011us}. This latter approach is a form of effective field theory including renormalization group (RG) flow. There is no clear controlled power counting in such effective field theories but they nevertheless provide some insight into the dynamics and quantitative behaviour of the gauge dynamics. The effects of magnetic field in gauge theories with known duals has been studied in \cite{Johnson1,Erdmenger:2007bn,Evans:2010iy,Callebaut:2013ria,Jokela:2013qya}. \cite{Rougemont:2015oea} has explored the effect of magnetic field on the deconfinement transition associated in holography with the transition to a geometry with a black hole. AdS/QCD style models concentrate on the mesonic physics rather than gluonic physics and some explorations of magnetic field have been made in \cite{Callebaut:2013ria,Dudal:2015wfn}. Our work adds to this latter literature. 

Our effective holographic model is the Dynamic AdS/QCD model introduced in \cite{Alho:2013dka} which is based on the DBI action of a probe D7 brane \cite{Erdmenger:2007cm,Karch:2002sh,Grana:2001xn,Bertolini:2001qa,Kruczenski:2003be,Babington:2003vm}. The background space is AdS$_5$ so that there is a clear identification of the RG scale with the AdS radius, $\rho$. The chiral condensate $\s$ is identified with a scalar field in the bulk (from the top down intuition this can be thought of as a brane embedding). The QCD dynamics at $T=B=0$ is introduced by hand by giving the AdS scalar a radially dependent mass term so that the anomalous dimension, $\gamma$, of $\s$ matches that in QCD perturbation theory (naively extrapolated to the strongly coupled regime near $\Lambda_{QCD}$). When the mass runs through the Breitenlohner-Freedman (BF) bound \cite{Breitenlohner:1982jf} (at which point $\gamma=1$) the scalar becomes unstable and chiral condensation occurs\cite{Jarvinen:2011qe,Kutasov:2012uq,Alvares:2012kr}. Finite temperature can be introduced by replacing the AdS space with an AdS Schwarzschild black hole of the appropriate radius \cite{Witten}. Here a single phenomenological parameter \cite{Evans:2011eu} allows us to dial the order of the phase transition and we pick it to give an appropriate second order transition to match QCD \cite{Ejiri:2009,Aoki:2006}. As pointed out in \cite{Evans:2011eu} such models naturally predict that the meson melting transition \cite{Babington:2003vm,Apreda:2005yz,Mateos:2006nu,Hoyos:2006gb} occurs ahead of the chiral transition; in a top down picture, the flavour brane of the model smoothly encounters and then moves along the black hole horizon to reach the chirally symmetric phase. To this point the model has been tuned to match expectations in QCD. 

We next introduce the two lowest order terms that link the chiral condensate field to a background magnetic field and study their impact. The coefficients of these terms are not a priori fixed and it is not surprising that by picking  signs they can be made to favour or disfavour chiral condensation. The two terms have different temperature dependence so one can also play them off against each other to find regions of parameter space in which there is magnetic catalysis at low $T$ but inverse magnetic catalysis at high $T$. This confirms that the effective description allows the behaviour seen in the QCD lattice simulations which is a positive sign for the approach but perhaps not surprising given the freedom of the model.

\begin{figure}[]
 \centering 
{
\includegraphics[width=9cm]{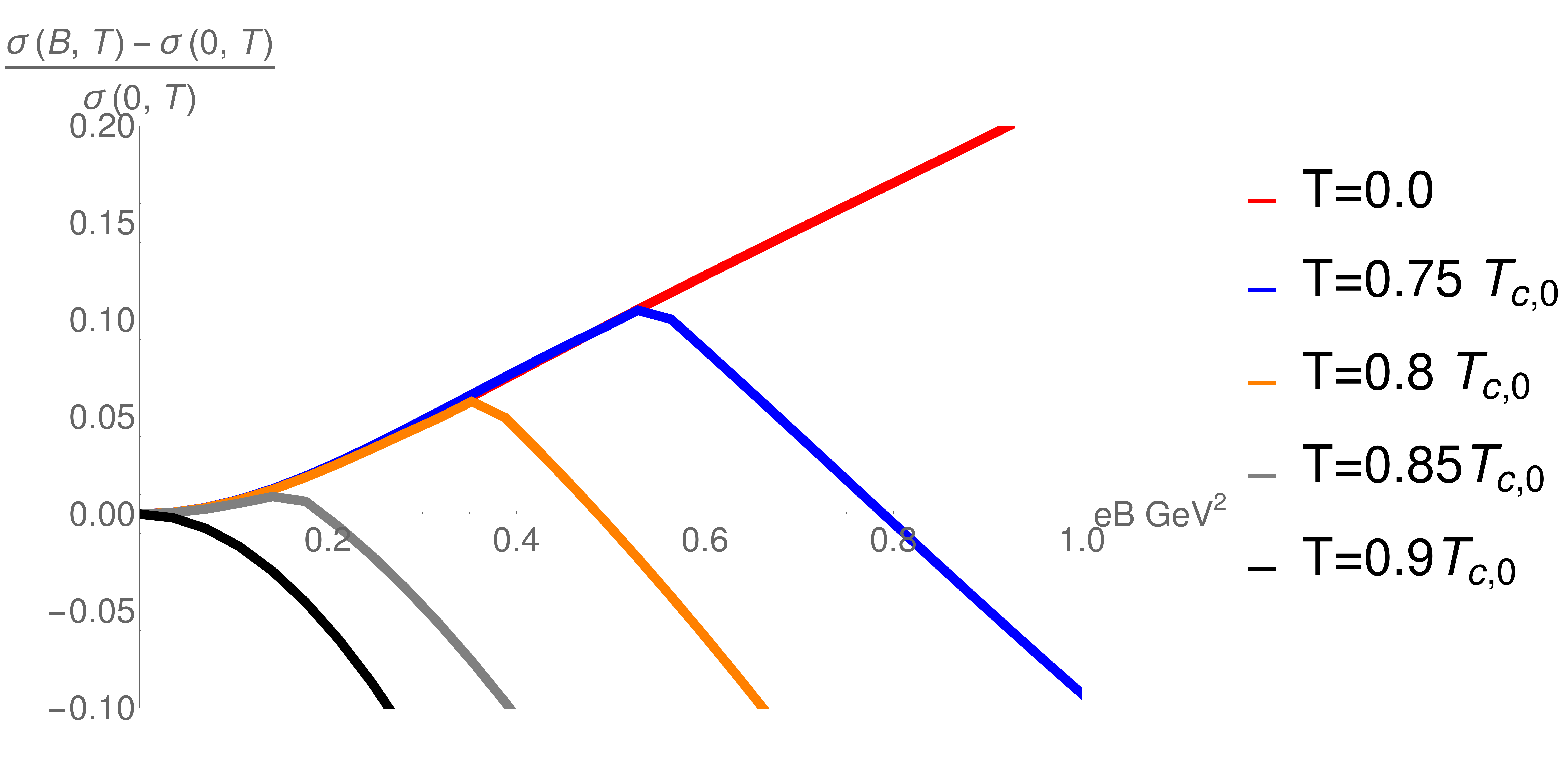}  }
  \caption{The holographic model's results for the change in the condensate of light quarks as a function of magnetic field strength over a range of temperatures for $N_c=N_f=3$. $T_{C0}$ is the thermal transition temperature at $B=0$ which is used to set our holographic energy scale (and is approximately 160 MeV according to the lattice simulations).
	For this plot the model parameters are taken as $\k = 0.05$ and $b=0.037$.}
            \label{fig:svB}
\end{figure}

The model offers some more interesting insights though. It turns out that is possible to reproduce the QCD behaviour with just a single one of the two terms. The reason is that the term produces magnetic catalysis in the low $T$ phase where the brane embedding lies off the black hole but inverse catalysis for the meson melted phase where the brane lies on the black hole. In the intermediate regime in the model $\s(B,T)$ exhibits non-monotonicity, enhancing the condensate for small $B$ but suppressing it at larger values. A summary of the typical behaviour we find is shown in Fig  \ref{fig:svB} for comparison with Fig \ref{fig:lattice}. The similarity in the generic behaviour is quite striking although the turn over is sharp in the holographic model since it is associated with the second order meson melting phase transition. The sharpness could be an artefact of large N though where the black hole description of temperature is very sharp (the horizon). We believe the holographic model sheds interesting light on the QCD results though.

Adjustment of the UV boundary conditions on the bulk field describing $\s$ allows the study of heavier quarks also. Generically these are associated with embeddings that do not touch the black hole horizon and we see magnetic catalysis for such configurations persist to larger B. One should be careful though not to extrapolate the results to too large $m$ or $B$ since the holographic framework is presumably unreliable  when the key physics is happening in the asymptotically free, weakly coupled regime.  In this sense larger quark masses are associated with just catalysis in the allowed regime of validity.

\section{A Holographic Model of QCD}

Our model is sited in an AdS-space-time 
\begin{equation} ds^2 = \rho^2 dx_{3+1}^2 + {d\rho^2 \over \rho^2} \end{equation}
to give a clear interpretation of $\rho$ as the energy scale of the theory. The running of the theory will then be introduced through $\rho$ dependent dimensions and couplings for the fields in the AdS-bulk representing operators in the gauge theory \cite{Alho:2013dka}. 

For example, we introduce a dimension one field $L(\rho)$ that describes the quark condensate with action
\begin{equation} \label{simpleaction} S = \int d^4x d\rho ~ \rho^3 (\partial_\rho L)^2. \end{equation}
The equation of motion $\partial_\rho (\rho^3 \partial_\rho L)=0$ has solution $L= m + \s/\rho^2$ where $m$ is interpreted as the quark mass and $\s$ as the quark condensate (at $m=0$). In order for the QCD dynamics that triggers chiral condensation to be present, we include a running anomalous dimension for the quark condensate via a $\rho$ dependent mass term for $L$. Such an additional term to the action, for example,
\begin{equation} \Delta S = \int d^4x d\rho ~ \rho \Delta m^2(\rho) L^2, \hspace{1cm} \Delta m^2 = - 2 \gamma, \end{equation}
yields solutions for $L(\rho)$ at small, fixed $\gamma$ of \begin{equation}L \sim {m\over \rho^\gamma} + \frac{\s}{ \rho^{2-\gamma}}.\end{equation}
The quark condensate has dimension $3-\gamma$ and the quark mass dimension $1 + \gamma$ in agreement with the usual definition of the anomalous dimension.

To impose the QCD dynamics, which we stress we will include by ansatz, we use the perturbative QCD one loop result for $\gamma$
\begin{equation} \Delta m^2 = - 2 \gamma = -{3 (N_c^2-1) \over 2 N_c \pi} \alpha\end{equation}
 and the two loop running result for $\alpha$
\begin{equation} 
\mu { d \alpha \over d \mu} = - b_0 \alpha^2 - b_1 \alpha^3,
\end{equation}
where
\begin{equation} b_0 = {1 \over 6 \pi} (11 N_c - 2N_f), \end{equation}
and
\begin{equation} b_1 = {1 \over 24 \pi^2} \left(34 N_c^2 - 10 N_c N_f - 3 {N_c^2 -1 \over N_c} N_f \right) .\end{equation}
Asymptotic freedom is present provided $N_f < 11/2 N_c$. There is an IR fixed point with value
\begin{equation} \alpha_* = -b_0/b_1\,, \end{equation}
which rises to infinity at $N_f \sim 2.6 N_c$. Using the perturbative ansatz for the running is only valid in the far UV of the theory but it provides a sensible guess as to the form of the running in QCD. In fact for $N_c=N_f=3$, the main case we study here, the two loop contribution has only a small effect on the deep IR. Going to two loops does allow study of walking and conformal window dynamics\cite{Evans:2014nfa, Erdmenger:2014fxa}. We have explored that larger $N_f,N_c$ parameter space with $B$ field but the behaviours we find are very like those reported here for the $N_c=N_f=3$ QCD case so we do not present them here. 

Naively one would set $\mu = \rho$ to place the perturbative running in the holographic model. As one runs down from the UV, $\gamma$ eventually runs through 1 at the scale $\Lambda_{QCD}$ and the BF bound is violated for the scalar $L$. $L$ becomes unstable and develops a vev corresponding to chiral condensation. With $\mu = \rho$ this instability triggers an unbounded growth of $L$ in the IR. This unbounded behaviour should be cut off when $L \sim \Lambda_{QCD}$ - the simplest way to encode this is to set $\mu = \sqrt{\rho^2 + L^2}$ (and indeed this is the typical functional form in top down models of chiral symmetry breaking such as in the D3/D7 system \cite{Erdmenger:2007cm,Karch:2002sh,Grana:2001xn,Bertolini:2001qa,Kruczenski:2003be,Babington:2003vm}). We will therefore adopt this simple ansatz. 

Given a choice of the running of $\gamma$, the Euler-Lagrange equation for $L$ can be solved numerically. To achieve the chiral limit one requires $L \rightarrow 0$ as $\rho \rightarrow \infty$ so $m=0$. The IR boundary condition following from variation of the action is $\partial_\rho L=0$. We choose to impose this at the scale where $L=\rho$, corresponding to the on mass shell condition for the quark. If we were to extend the theory below this scale, we would have to include the decoupling of the quarks from the QCD running below that scale. This is not necessary to study the chiral condensation. 
We now have a model of the chiral dynamics of QCD that we can use to investigate the additional effects of temperature and magnetic field.

\section{Finite Temperature}

Finite temperature can be included in the model by replacing the background metric by AdS-Schwarzschild with the horizon at $r=r_H$ \cite{Witten},
\begin{equation}  ds^2 = \rho^2 (-f dt^2 + dx_{3}^2) + {d\rho^2 \over f \rho^2}, \hspace{1cm} f = 1 - {r_H^4 \over r^4},\end{equation} where $r_H$ is proportional to the temperature ($T=r_H/ \pi$). 
The action for $L$ becomes
\begin{equation} S = \int d^4x d\rho ~ \left(\rho^3 f (\partial_\rho L)^2 + \rho \Delta m^2(\rho) L^2 \right) \end{equation}
Again naively one would set $r=\rho$ but, as for the $\Delta m^2$ term discussed above, we want the field $L$ to decouple from the IR dynamics when it is large (a very heavy quark will be indifferent to a very low temperature) and so we instead adopt
\begin{equation} r = \sqrt{ \rho^2 + \kappa L^2}. \end{equation}

\begin{figure}[]
 \centering 
{\includegraphics[width=8.5cm]{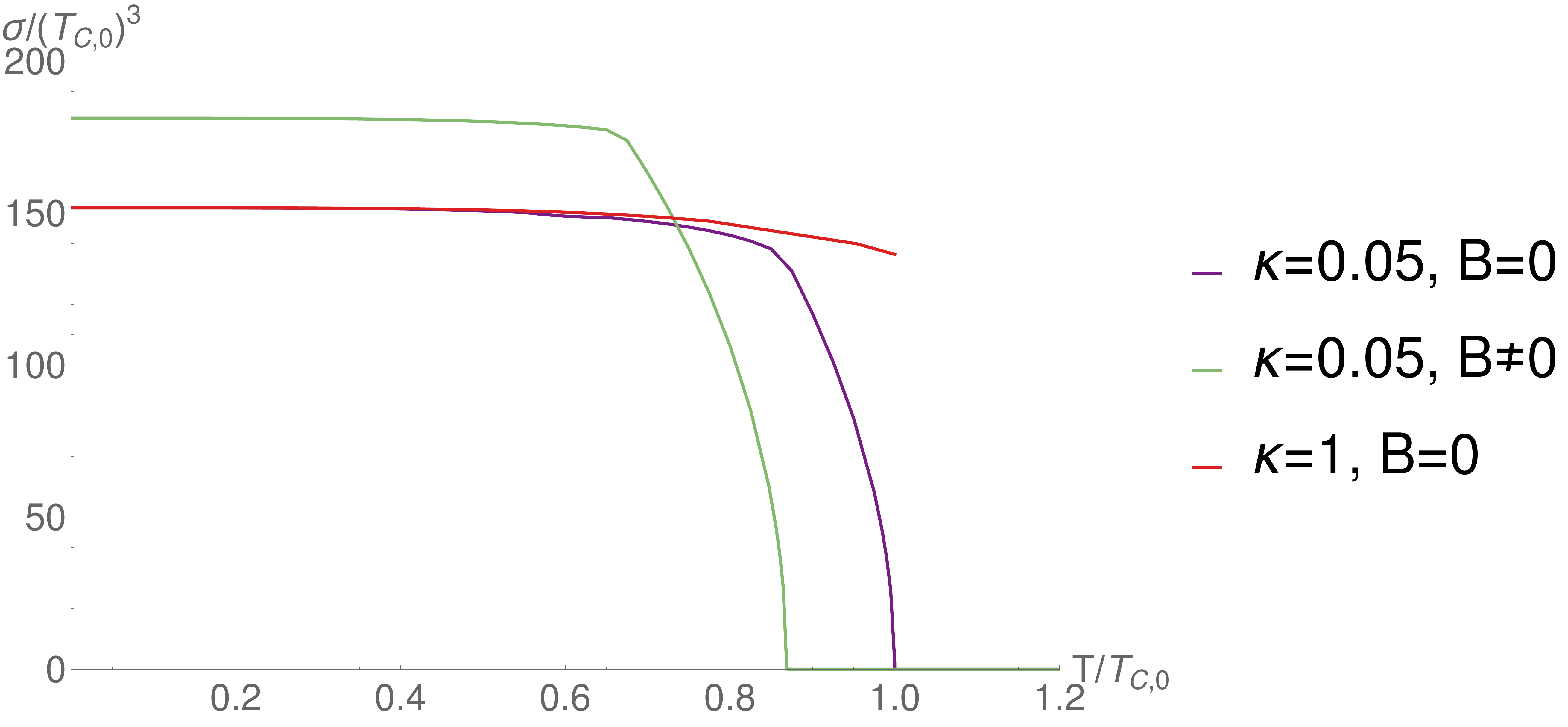}   }
  \caption{Thermal phase transitions in the holographic model with $N_c=N_f=3$. For values of the parameter $\k$ close to $\k=1$, a first order chiral transition is present. As the value of $\k$ is reduced and the black hole is deformed along the $L$-axis, the phase transition switches to becoming second order. The introduction of a background magnetic field can be seen to affect the value of the transition order parameter, $\s$. Here we show an example with magnetic catalysis at low temperature and inverse magnetic catalysis at higher temperatures, a phenomenon which reduces the critical temperature, $T_c(B)<T_c(0)$.  }
            \label{fig:svT}
\end{figure}
\begin{figure}[]
 \centering 
{\includegraphics[width=9.5cm]{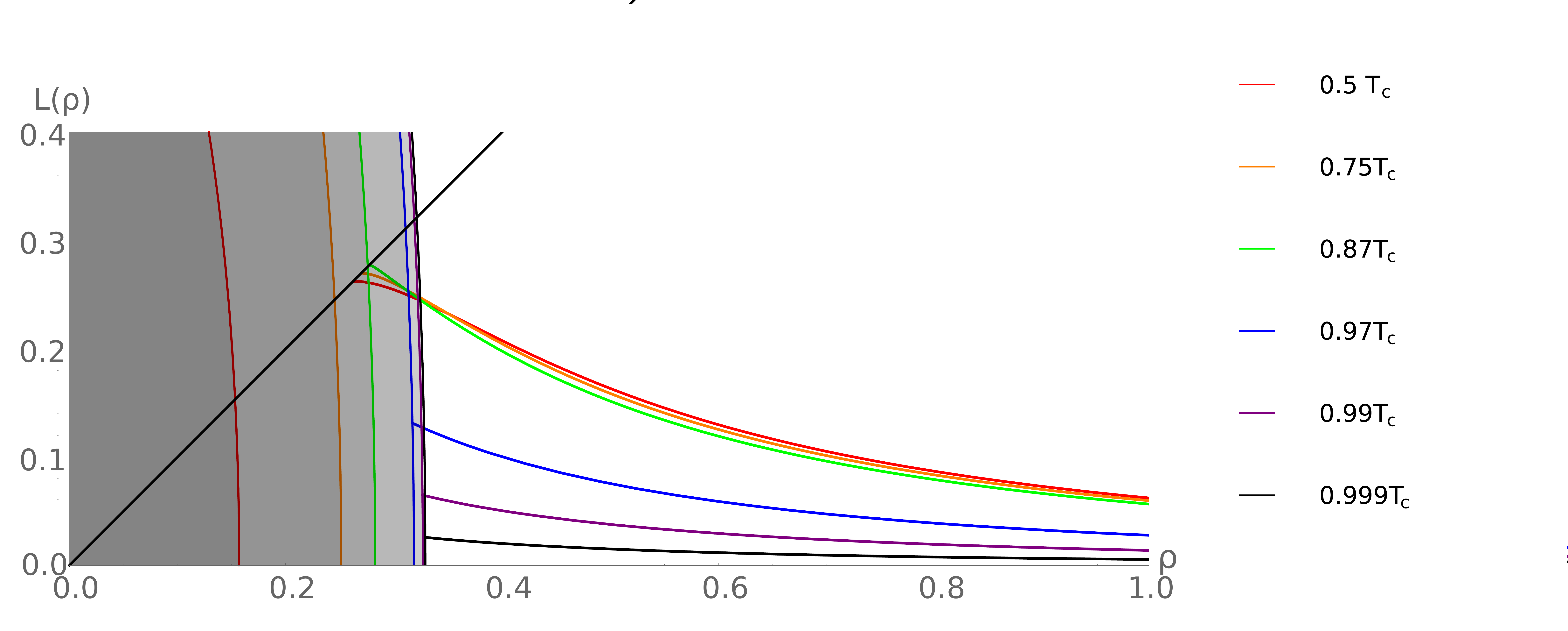}   }
  \caption{Plot showing the chiral embeddings (at $B=0$) for a range of temperatures. Each embedding is coloured to match the black hole horizon pertaining to the relevant temperature. The second order nature of the transition is evident; the embedding smoothly transforms into the flat $L=0$ embedding at the critical temperature, $T_c$.}
            \label{fig:Lvrho}
\end{figure}

Most simply one would set $\kappa=1$.  Here one again seeks numerical solutions of the Euler-Lagrange equations subject to $L \rightarrow 0$ as $\rho \rightarrow \infty$ to describe massless quarks. In the IR one chooses either $\partial_\rho L=0$ at the on-mass shell condition or for the end point of the flow to lie on the black hole.  As shown in Fig \ref{fig:svT} the transition for $\kappa=1$ is first order. The main signal of the first order transition is that in the transition region there is an horizon ending solution which at low temperatures emerges from the L=0 embedding and then moves up to join the solution off the black hole at higher temperature. This extra solution corresponds to the local maximum of the effective potential between the chiral symmetry breaking and the chirally symmetric solutions. 

As pointed out in  \cite{Evans:2011eu}, lower values of $\kappa$ turn the transition second order. In the second order case there is again an horizon ending solution for $L$ but it now emerges at low T from the chiral symmetry breaking solution and then moves to merge with the $L=0$ embedding at higher temperature.  In Fig \ref{fig:svT}, we plot the chiral condensate, $\s$, against $T$ for  $\kappa=0.05$. In Fig \ref{fig:Lvrho}, we show the explicit second-order behaviour at $\k=0.05$ by plotting the  embeddings of the scalar $L$ in the chiral limit (i.e. $L(\rho\to\infty)=0$) as the temperature is increased towards the critical value. At $\rho=1$ it is already evident that the value of the condensate, proportional to the gradient of the embedding $\partial_\rho L|_{\rho\to\infty}$, decreases with increasing temperature. Moreover, as we approach the critical temperature, the angle subtended by the arc of the black-hole horizon between $L(\rho)$ and $L=0$ decreases smoothly to zero at $T=T_c$. Above this value only the flat, chirally symmetric $L=0$ embedding exists.

It's important to stress the physics of the two continuous transitions shown in Fig \ref{fig:Lvrho}. At low temperatures the embedding lies off the black hole horizon and small fluctuations about the embedding are associated with mesonic modes \cite{Kruczenski:2003be}. They are stable in this phase. When the embedding moves on to the black hole the mesonic fluctuations become replaced by quasi-normal modes that describe unstable plasma fluctuations \cite{Hoyos:2006gb} - the mesons of the theory have melted. The configuration then continues to evolve with $T$ until a flat embedding is reached and chiral symmetry is restored. Clearly in a second order transition these two transitions must be separate and the meson melting must occur earlier. 

In the effective description of the model we view $\kappa$ as a parameter one must adjust to correctly reproduce the expected phase structure at a given $N_c,N_f$. To represent QCD we will choose the second order behaviour and $\kappa=0.05$.  We will use the value of $r_H$ at which the transition occurs at $B=0$ to set the scale of the radial energy direction, $\rho$, in the holographic model. We set $T_{C0} = 160$MeV.

\section{Magnetic Field}

Background U(1)$_B$ electromagnetic fields are introduced into AdS/CFT via sources for the operator $\bar{q} \gamma^\mu q$\cite{Johnson1,Erdmenger:2007bn,Evans:2010iy,Callebaut:2013ria,Jokela:2013qya}. These operators are described by a bulk, massless U(1) gauge field. The quark condensate has no baryon number charge so interactions will be products between the field $L$ plus its derivative $\partial_\rho L$ and $F^2$. The leading two terms in an expansion in $L$ are
\begin{equation} \label{aandb}\Delta S = \int d^4x d\rho ~ \left(a  \rho F^2 L^2 + b  \rho^3 f F^2 (\partial_\rho L)^2\right).\end{equation}
In the case of a fixed external magnetic field, including the metric factors, $F^2=B^2/\rho^4$ and we will treat $a$ and $b$ as phenomenological parameters. The $a$ term is then a direct $B$ dependent contribution to the running of the $L$ mass or anomalous dimension of the quark condensate. Clearly choice of the sign of $a$ can favour or disfavour chiral condensation by effecting where the BF bound is violated. The $b$ term, again depending on the sign, either favours or disfavours curvature in the $L$ profile which again encourages or discourages L to take up a profile away from $L=0$ (the chirally symmetric state).   Note that the magnetic field now enters into the action in the combinations $aB^2$ and $b B^2$ so it is possible by choice of the magnitude of $a$ and $b$ to move the scale of effects in $B$.

Interestingly, only the second term has temperature dependence when one naively inserts the metric factors (from the $\rho$ index contraction in $(\partial_\rho L)^2$). This term decreases as one approaches the black hole horizon. One can hope to play these two terms off against each other. At zero temperature the second term might dominate and favour chiral condensation. At finite temperature though it will be less favoured and the first term might take over suppressing chiral condensation. This is our initial strategy to realize the observed pattern of catalysis and inverse catalysis with temperature. 

The numerical analysis is again to find the solutions for $L$ at each value of $T$ and $B$ for our chosen values of $a,b$ and $\kappa$, $N_f$=$N_c$=3. For $\kappa$ of order one the thermal transition is first order. The embedding profile for $L$ jumps from a solution off the black hole to the flat embedding ending on the horizon. The transition is driven by the black hole ``eating" the $L=0$ configuration until its action is less than the chiral symmetry breaking embedding. For this reason the chirally symmetric low $T$ phase is fairly insensitive to the temperature and it is very hard to engineer $T$ dependent behaviour. The only shifts from magnetic catalysis to inverse catalysis that we can find occur when  the $a$ and $b$ terms are so finely tuned that they have negligible net effect at $T=0$. The catalysis effect is well below a percent. We conclude that the QCD behaviour with $B,T$ is a result of the second order transition behaviour.  

Hence we turn to $\kappa=0.05$ as an example of a model with a second order transition. For each point in the $a,b$ plane we can plot the condensate against $T$ at non-zero B. In all cases the transitions are second order.  An example curve is shown in Fig \ref{fig:svT} for a case where the condensate is enhanced at small $T$ but suppressed at $T \simeq T_c$. 

\begin{figure}[]
 \centering 
{
\includegraphics[width=8.5cm]{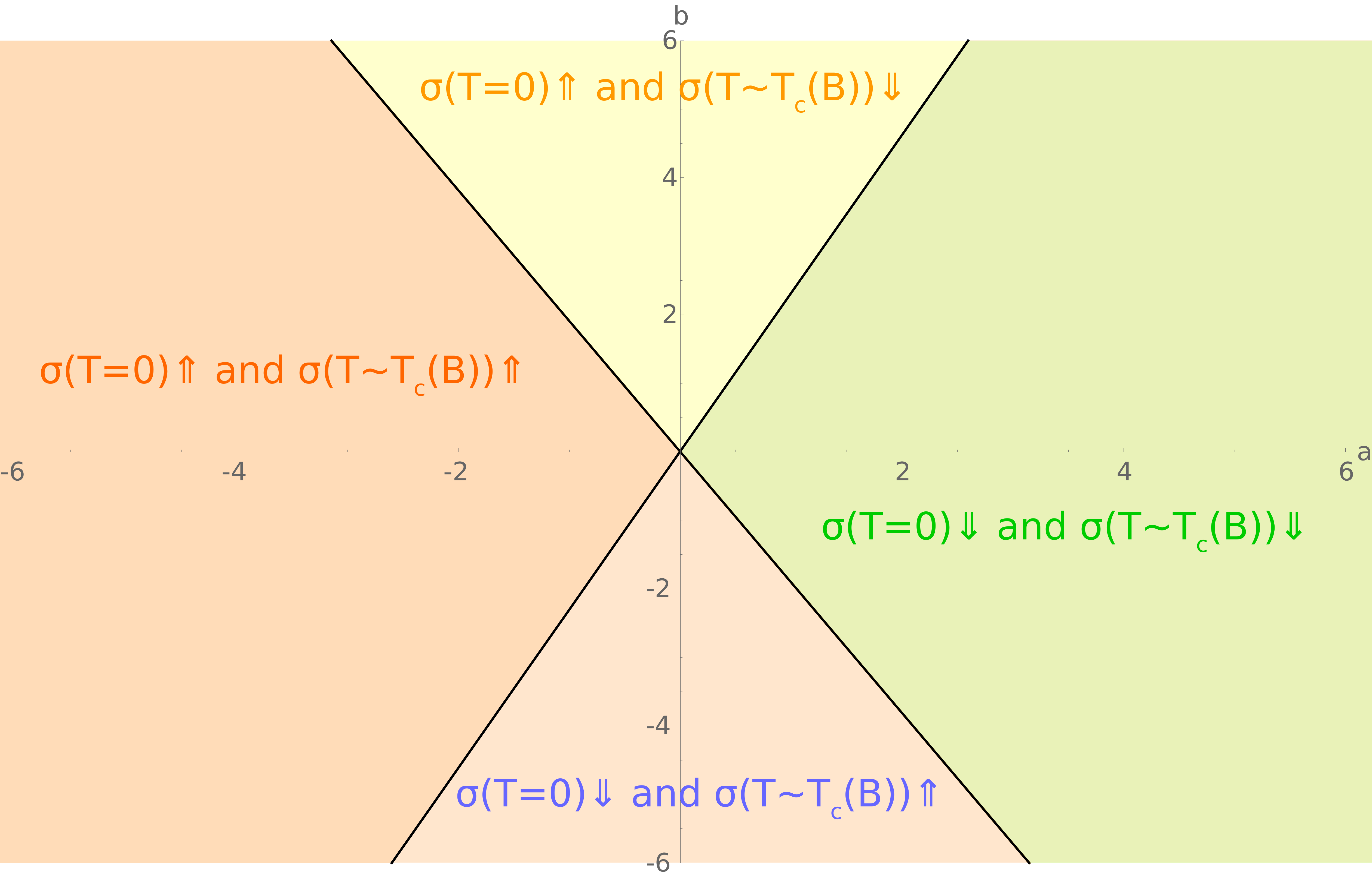}  }
  \caption{ The phase-structure of the holographic model in terms of the phenomenological parameters $a$ and $b$. The $a-b$ plane can be dissected into four sectors wherein the condensate is affected differently with temperature and an external magnetic field.}
            \label{fig:avb}
\end{figure}

In Fig \ref{fig:avb}, we show the phase structure of the model in terms of the phenomenological parameters $a$ and $b$. This $a-b$ phase space comprises four different sectors; a region in which the chiral condensate, $\s$, is \emph{always} enhanced relative to no external magnetic field, a region in which $\s$ is \emph{always} suppressed relative to no external magnetic field and two regions where it is either enhanced at low temperatures and suppressed at high temperatures or vice versa. 

\begin{figure}[]
 \centering 
{
\includegraphics[width=8.5cm]{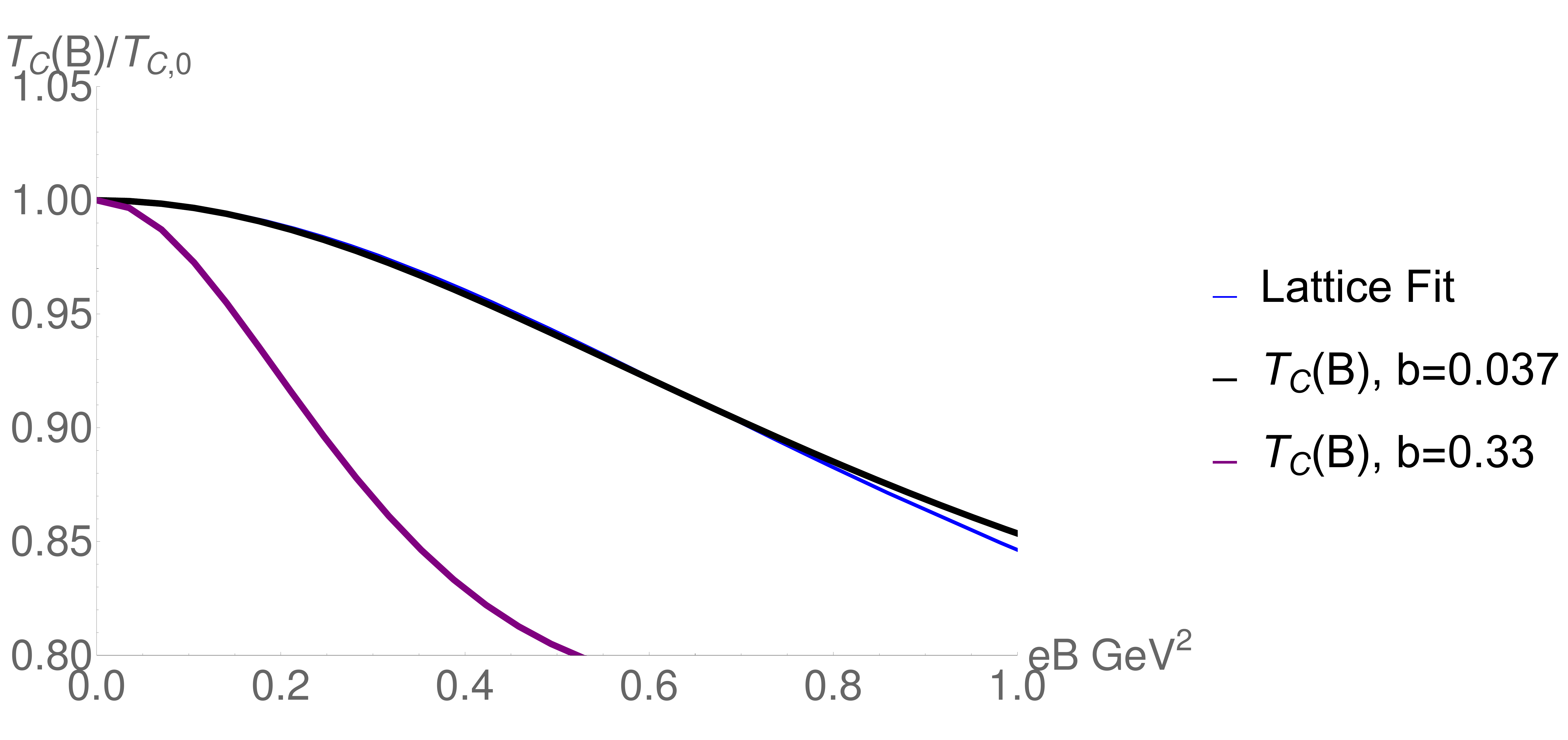}   }
  \caption{ Plots of the critical temperature against $eB$, $T_{C0}=160$MeV. We show the best fit lattice data taken from \cite{Endrodi:2015oba} and the holographic models best fit to that data ($\k=0.05, a=0, b=0.037$). We also show the holographic models prediction for another value of $b=0.33$ - the model depends on the quantity $b B^2$ so the $eB$ axis is simply rescaled by this change.   }
            \label{fig:TvB}
\end{figure}

The value of the critical temperature of the chiral phase transition is dependent on how the external magnetic field affects the value of the condensate. If at high temperatures the value of the condensate is suppressed due to inverse magnetic catalysis, the value of the critical temperature is reduced, see an example in Fig \ref{fig:TvB}, or if enhanced at high temperatures the critical temperature is increased.

The Lattice QCD data from \cite{Bali:2012zg}, shown in Fig \ref{fig:lattice}, indicates a non-trivial relationship between the chiral condensate and the strength of the external magnetic field. At low temperatures, magnetic catalysis of the condensate is apparent and at temperatures approaching the critical value, there is a suppression of the condensate with $B$ (or inverse magnetic catalysis). This, first of all, points us to working in the top centre quadrant of the $a-b$ plane seen in Fig \ref{fig:avb}. It is encouraging that the holographic model can incorporate the QCD behaviour although the freedom of the $a,b$ parameter space suggested it should be possible.  

A further interesting feature of the lattice plot  is that for a narrow range of temperatures approaching the critical value, $\s(B)$ behaves non-monotonically, indicating magnetic catalysis for small values of the magnetic field but as the strength of the external $B$-field is increased, the field catalyses a suppression - we will refer to this intermediate behaviour as the `cross-over' regime. One question we could ask of our model is whether or not this cross-over behaviour can be obtained if one were to select values of the phenomenological parameters $a$ and $b$ to be inside the appropriate sector of the $a-b$ plane. 

\begin{figure}[]
 \centering 
{\includegraphics[width=9.cm]{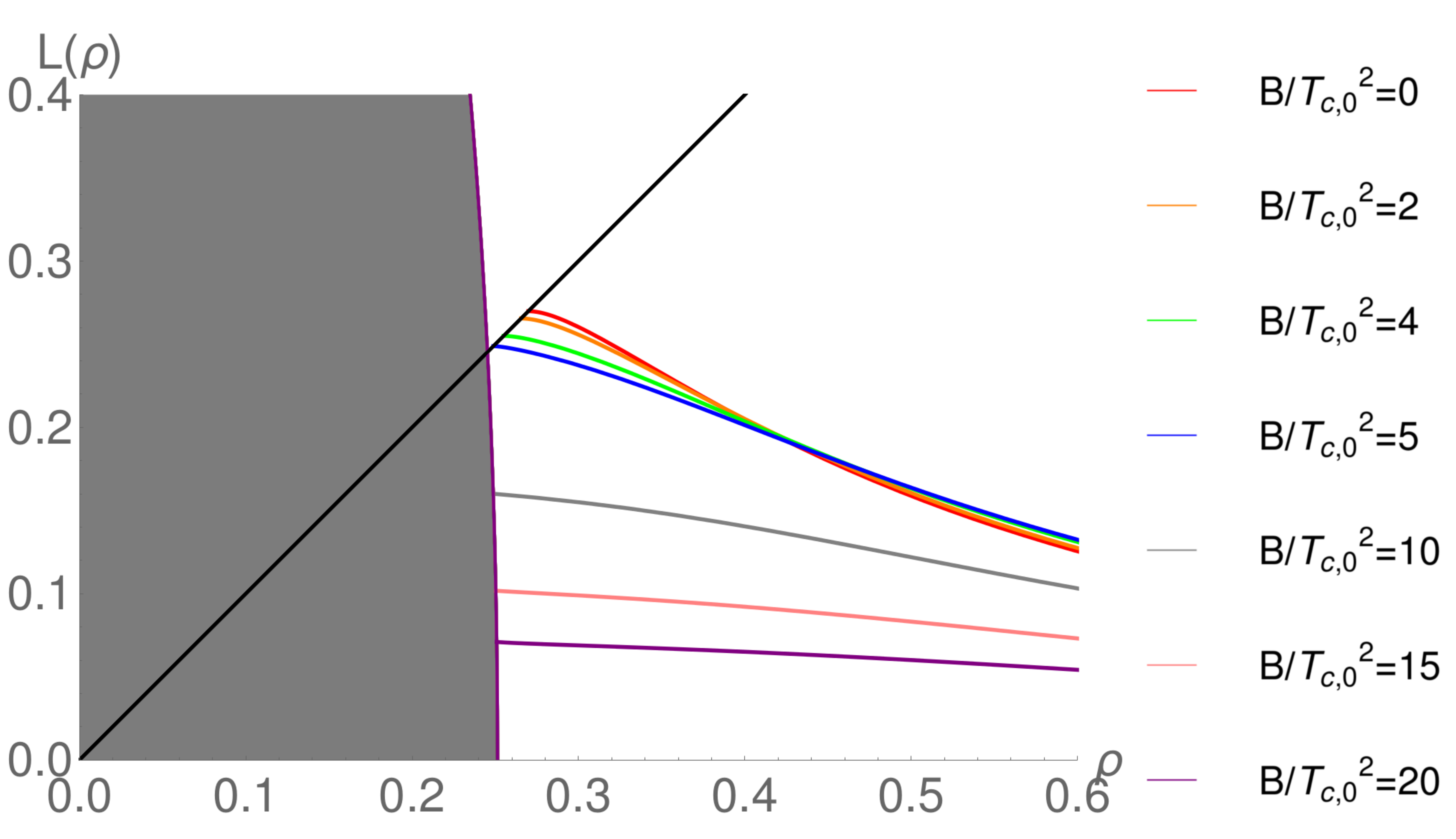}   }
  \caption{Plot showing the chiral embeddings at $T=0.75 T_{C0}$  for a range of magnetic field. Increasing $B$ moves the embedding towards and on to the horizon but initially also increases the condensate value.}
            \label{fig:Lvrho2}
\end{figure}

The key observation that allows us to achieve this cross-over behaviour in the holographic model is to notice that  the appropriate quadrant in Fig \ref{fig:avb} contains the $b$ axis where $a=0$. Intriguingly the $b$ term alone can generate magnetic catalysis at low temperatures but inverse catalysis at higher temperatures. Further exploration shows that the reason is that the term acts differently on black hole embeddings and non-black hole embeddings. In Fig \ref{fig:Lvrho2} we show the effect of $B$ on the embeddings at an intermediate temperature $T=0.75 T_c$. At zero $B$ we are still in the phase before the mesons have melted. As $B$ rises in the theory with just the $b$ interaction term, derivatives are encouraged in the UV but not close to the horizon where the $b$ term dies due to its $T$ dependence. The result is that in the IR the B field moves the embedding towards a melted phase whilst the UV condensate grows. Once the embedding is brought onto the black hole further $B$ moves the embedding down the horizon and then decreases the UV condensate. We did not deliberately engineer this behaviour but it does match the observed lattice results. 

 One can now survey the $a,b,\kappa$ space for the best fit to the QCD behaviour. We have found a decent fit to the QCD behaviour when we take $\k =0.05$ and $a=0$. To fit $b$ we have used the lattice fit in \cite{Endrodi:2015oba} for the temperature dependence of the critical temperature in the theory. There they fit the form
\begin{equation} {T_C(eB) \over T_{C0} }= {1 + \alpha (e B)^2 \over 1 + \beta (eB)^2} \end{equation}
where $e^2/4 \pi = 1/137$. The lattice results find central values, from fitting to the the light quark condensate, $\alpha=0.54$ and $\beta=0.82$.
In Fig \ref{fig:TvB} we show our fit to this data for $b$ - the lattice and holographic models can be made to lie very close to each other when $b=0.037$ - the holographic model best fits the functional form with $\alpha=0.78$ and $\beta=1.08$. 
Now, with all parameters fixed, we can plot the fractional change in the condensate against $eB$ at different $T$ as shown in Fig \ref{fig:svB}. We see the enhancement of the condensate at zero temperature but a suppression near the critical value. Of course it should be reiterated that without the lattice data already in place, we do not know \emph{a priori} which values of $a$ and $b$ should be chosen to best fit QCD. Having chosen these appropriate values for the parameters $a$ and $b$, it is no surprise that we reproduce the expected enhancement and suppression of $\s$ at low and high temperatures respectively. More remarkable however, as we have discussed, we also find the cross-over regime needed by Fig \ref{fig:lattice}. For intermediate temperatures, a transition occurs at some value of $B$ at which the condensate switches from increasing to decreasing with an increasing magnetic field strength. The turn over point of this transition can be identified in the holographic model as the value of the magnetic field at which the chiral embedding switches from being off the black hole to being a solution ending on the black hole i.e. the meson melting phase transition. The match between Fig \ref{fig:lattice} and Fig \ref{fig:svB} is not perfect - the holographic model has less catalysis at low $T$ and too much inverse catalysis at higher $T$ but the general structure is similar. We hope that we have learnt from the holographic model that the meson melting behaviour is key to the structure of the transitions seen with $B$.  

\section{Quark Mass}

\begin{figure}[]
 \centering 
{\includegraphics[width=9.cm]{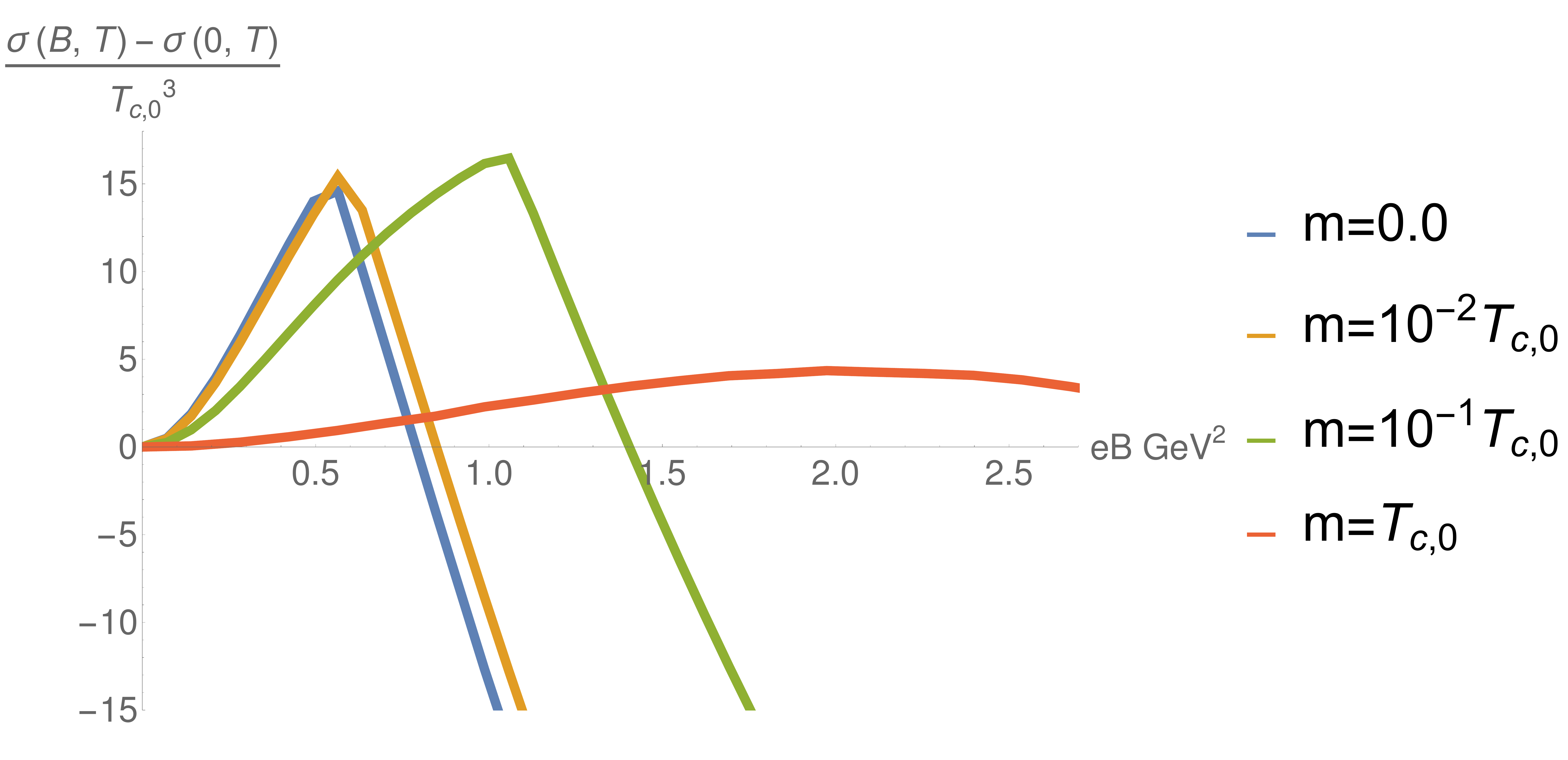}   }
  \caption{Plot showing the change in the chiral condensate as a function of B for different quark masses. The parameters are those used to make Fig  \ref{fig:avb}.}
            \label{mass}
\end{figure}

It is straightforward to include quark mass into the analysis. The asymptotic value of the field $L$ is simply the quark mass (as discussed under (\ref{simpleaction})) and we can set it to some finite value at a large UV scale. To compute the quark condensate one should determine the derivative with respect to $m$ of the Lagrangian evaluated on the vacuum solution. Since the relevant action terms are of the form $L^2$ there is a leading quadratically divergent piece $m^2 \Lambda_{UV}^2$ which gives a divergent contribution to the condensate which we subtact. The cross term takes the form $m \sigma$ and therefore $\sigma$ remains as the leading finite condensate term.  We have repeated our analysis for the parameters used to draw Fig \ref{fig:svB} but with varying UV quark mass. We show the variation in the quark condensate with $B$ in Fig \ref{mass}. Raising the mass moves the B at which inverse catalysis takes over from catalysis to higher values. Since the effective theory does not apply at B field values that begin to probe the asymptotically free regime, and since perturbative analysis suggest only magnetic catalysis, this suggests that our results might smoothly move to an absence of catalysis at large $m$. It is indeed found on the lattice that inverse catalysis results only for small quark masses. 

\section{Conclusion}

Lattice data \cite{Bali:2012zg} indicates that at low temperatures the QCD condensate is enhanced in the presence of an external magnetic field, whilst close to the transition temperature, the magnetic field strength induces inverse magnetic catalysis of the condensate, suppressing its value.

We have presented an AdS/QCD model with the running of the anomalous dimension of $\langle\bar{q}q\rangle$ inserted by hand to match a particular theory - in our case $N_c=N_f=3$, chiral QCD. The field holographic to the quark bilinear, $L$, is adjusted so that is has a mass dependent on the radial coordiate of the bulk AdS-space. This is equivalent to  introducing an anomalous dimension in the field theory, in essence making the field theory dynamical. Temperature can be introduced through a black hole geometry and we included a phenomenological parameter $\kappa$  \cite{Evans:2011eu} that allows us to engineer a second order chiral phase transition with temperature as observed in lattice QCD. We set the holographic energy scale by the position of the $B=0$ phase transition at $T_{C0}=160$MeV. 

Magnetic field enters the holographic model through two terms linking the condensate and a bulk baryon number gauge field (\ref{aandb}). We have shown (see Fig  \ref{fig:avb}) that, for the zero quark mass theory, magnetic catalysis at low T and inverse catalysis at high T can be achieved in the parameter space of the model. The model also provides for free an explanation of the non-monotonic behaviour of the condensate with $B$ at intermediate $T$ seen on the lattice - compare the lattice data in Fig \ref{fig:lattice} and the holographic model's equivalent plot in  Fig \ref{fig:svB}. In the holographic model the turn over is associated with a second order meson melting transition which occurs at a lower $T$ than the chiral transition. As the quark mass is raised form zero, the scale in B at which inverse catalysis sets in grows. Since the model is not valid at too large B this suggests the physics of large mass quarks is dominated by catalysis rather than inverse catalysis. This behaviour may be useful for understanding the observed lattice QCD data. \bigskip

\noindent {\bf Acknowledgements:} NE and MS are grateful for the support of STFC. CM thanks CAPES (Proc. 9397/2014-0).
We are grateful to Gunnar Bali, Gergely Endrodi and  Falk Bruckmann for their comments and encouraging us to fit to the lattice data.

\end{document}